\documentstyle[prb,aps,floats,epsf]{revtex}
\begin{document}
\draft 
\wideabs{
\title{Electronic structure of spinel-type LiV$_2$O$_4$}
\author{J.~Matsuno, A.~Fujimori and L.~F.~Mattheiss}
\address{Department of Physics, University of Tokyo, Bunkyo-ku, Tokyo
 113-0033, Japan}
\date{\today} 
\maketitle

\begin{abstract}
 The band structure of the cubic spinel compound LiV$_2$O$_4$, which
 has been reported recently to show heavy Fermion behavior, has been
 calculated within the local-density approximation using a
 full-potential version of the linear augmented-plane-wave method.
 The results show that partially-filled V 3$d$ bands are located
 about 1.9 eV above the O 2$p$ bands and the V 3$d$ bands are split
 into a lower partially-filled $t_{2g}$ complex and an upper
 unoccupied $e_{g}$ manifold. The fact that the conduction electrons
 originate solely from the $t_{2g}$ bands suggests that the
 mechanism for the mass enhancement in this system is different from
 that in the 4$f$ heavy Fermion systems, where these effects are
 attributed to the hybridization between the localized 4$f$ levels
 and itinerant $spd$ bands.
\end{abstract}
\pacs{PACS numbers: 71.27.+a, 71.20.-b, 71.15.Mb} 
}

%\vskip1pc]
\narrowtext 

The recent discovery of heavy Fermion (HF) behavior in LiV$_2$O$_4$ by
Kondo {\it et al.}~\cite{Kondo} has significant importance because
this is the first $d$-electron system that shows HF characteristics, a
phenomenon that has previously been observed only in $f$-electron
systems. Kondo {\it et al.} have reported a large electronic heat
coefficient of $\gamma \approx$ 0.42 J/mol K$^2$ and a crossover with
decreasing temperature $T$ from local moment to renormalized
Fermi-liquid behavior.~\cite{YMn2} Recently Takagi {\it et al.} have
reported that the electrical resistivity $\rho$ of single crystals
exhibits a $T^2$ temperature dependence $\rho = \rho_0 + AT^2$
(Ref.~\onlinecite{Takagi}) with an enormous $A$, which is another HF
characteristic. The Curie-Weiss law at high temperatures implies that
each V ion has a local moment and their coupling is
antiferromagnetic~\cite{Kessler} although no magnetic ordering is
observed down to 0.02 K.~\cite{Kondo}

The purpose of the present study is to determine from first principles
the electronic band properties of LiV$_2$O$_4$, with particular
emphasis on features near the Fermi level $E_F$. This provides a
reference framework for evaluating the magnitude of the heavy Fermion
mass-enhancement effects in this material. In addition, a simple
tight-binding model~\cite{TB} which captures the essential features of
the LiV$_2$O$_4$ electronic structure is presented. In heavy Fermion
systems, the enhanced electron mass manifests itself in terms of an
exceptionally large value of the density of quasi-particles at $E_F$.
Experimentally, this high density is reflected in the large
specific-heat $\gamma$ and the large spin susceptibility $\chi^{\rm
  spin}$ which is nearly $T$-independent. The calculated band DOS at
$E_F$ $D(E_F)$ can be compared directly with the experimental results,
and the ratio of the experimental and calculated band $D(E_F)$ values
then provides a direct measure of the enhancement factors.

\begin{figure}
\epsfxsize=80mm \centerline{\epsfbox{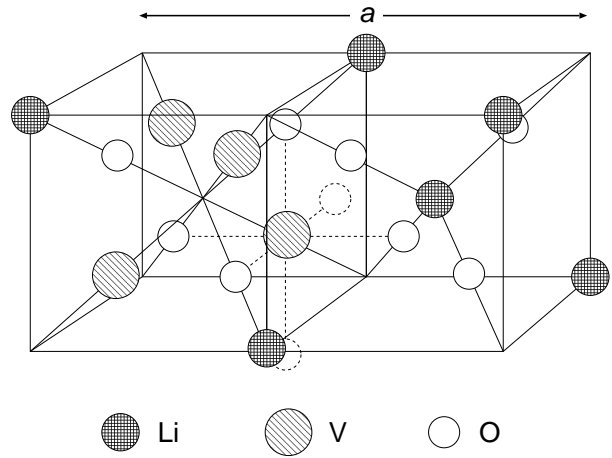}} \vspace{0.5pc}
\caption{A portion of the unit cell for the spinel-type LiV$_2$O$_4$,
 where $a$ is the lattice parameter for the face-centered-cubic
 Bravais lattice. (See text)} 
\label{structure} 
\end{figure}

LiV$_2$O$_4$ forms with the well known cubic spinel structure in which
the Li ions are tetrahedrally coordinated by oxygens while the V sites
are surrounded by a slightly-distorted octahedral array of oxygens.
The spinel structure features a face-centered-cubic Bravais lattice
and a nonsymmorphic space group ($Fd3m$) which is identical with that
of the diamond structure. The primitive unit cell contains two
LiV$_2$O$_4$ formula units (14 atoms). One can construct the spinel
structure by an alternate stacking along [100]-type directions of the
two different kinds of cubes shown in Fig.~\ref{structure}. The
LiV$_2$ substructure is the same as the $C15$ structure $AB_2$, where
the local moments at the $B$ sites are highly frustrated.~\cite{YMn2}
The observed lattice constant of the LiV$_2$O$_4$ is 8.22672 \AA\/ at
4 K.~\cite{Chmaissem} The eight oxygen atoms in the primitive cell are
situated at the 32$e$-type sites, at positions which are determined by
the internal-position parameter $x=0.2611$. The VO$_6$ octahedra are
trigonally distorted in the spinel structure unless the parameter $x$
is equal to the ``ideal'' value 0.25.

The present band-structure calculations for LiV$_2$O$_4$ have been
carried out in the local-density approximation (LDA) using a
full-potential, scalar-relativistic implementation~\cite{LFM} of the
linear augmented-plane-wave (LAPW) method.~\cite{Andersen} The LAPW
basis has included plane waves with a 14-Ry cutoff ($\sim$60
LAPW's/atom) and spherical-harmonic terms up to $l=6$ inside the
muffin tins. The crystalline charge density and potential have been
expanded using $\sim$7300 plane waves (60 Ry cutoff) in the
interstitial region and lattice harmonics with $l_{\rm max}=6$ inside
the muffin-tin spheres ($R_{\rm Li} \approx 1.98$ a.u., $R_{\rm V}
\approx 2.16$ a.u., $R_{\rm O} \approx 1.54$ a.u.). Brillouin-zone
(BZ) integrations have utilized a ten-point {\bf k} sample in the 1/48
irreducible wedge. Exchange and correlation effects have been treated
via the Wigner interpolation formula.~\cite{Wigner} The atomic
Li($2s^1$), V($3d^44s^1$) and O($2s^22p^4$) states were treated as
valence electrons in this study whereas the more tightly bound levels
were included via a frozen-core approximation.

\begin{figure}
\epsfxsize=80mm \centerline{\epsfbox{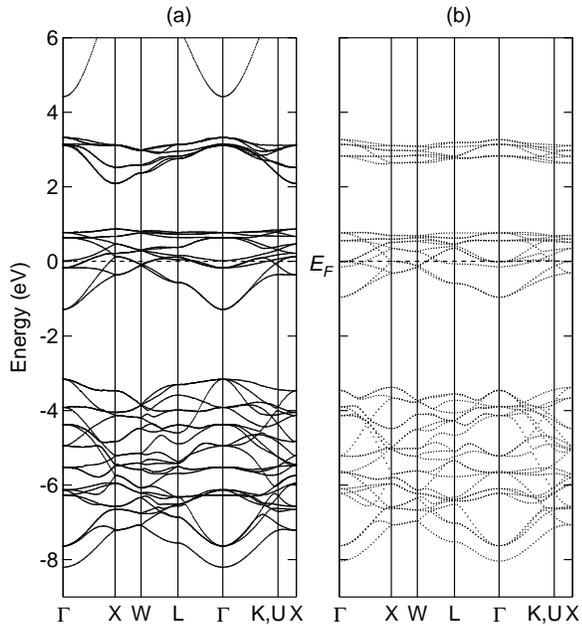}} \vspace{0.5pc}
\caption{Comparison of the LAPW (a) and tight-binding (b) energy-band
 results for LiV$_2$O$_4$.}
\label{band} 
\end{figure}

The results of the present LAPW calculations for LiV$_2$O$_4$ are
plotted along selected BZ symmetry lines in Fig.~\ref{band}(a). The O
2$s$ states, which are not shown, form narrow ($\sim$1 eV) bands and
are situated about 19 eV below the Fermi level $E_F$. The lowest band
complex shown in this figure evolves from the O 2$p$ states: it
contains 24 bands, and has an overall width of $\sim$4.8 eV. The 20 V
3$d$ bands lie above the O 2$p$ bands, separated by a 1.9-eV gap. A
broad plane-wave conduction band evolves from the $\Gamma$ point above
4.4 eV. These general features are quite similar to those exhibited by
previous results for the closely related isostructural compound
LiTi$_2$O$_4$.~\cite{Massidda,Satpathy}

The octahedral crystal field at the V sites in the spinel structure
splits the 20 V 3$d$ bands into twelve partially-filled $t_{2g}$ bands
and eight $e_{g}$ states. The Fermi level lies within the $t_{2g}$
complex, and thus the transport properties of LiV$_2$O$_4$ are solely
associated with $t_{2g}$ bands. The formal valence of vanadium is
3.5+, yielding exactly six valence electrons per cell in a perfectly
stoichiometric material. The point symmetry at the V sites is
$D_{3d}$. In the fully localized limit, this allows the cubic $t_{2g}$
crystal-field-type states to be split into $a_{1g}$ and $e_{g}$
levels. As discussed below, tight-binding estimates of this splitting
show that it is quite small ($\sim$0.14 eV) in LiV$_2$O$_4$, about 6
\% of the octahedral $e_{g}$-$t_{2g}$ splitting ($\sim$2.5 eV). The
0.14-eV difference between the $a_{1g}$ and $e_{g}$ orbital energies
is small compared to the overall $t_{2g}$ bandwidth ($\sim$2.2 eV).
As a result, this splitting of the $a_{1g}$ and $e_{g}$ orbital
energies does not produce effects which are readily discernible in the
band-dispersion curves for LiV$_2$O$_4$.

In order to provide a convenient starting point for future
investigations of electron-correlation effects in the present
LiV$_2$O$_4$ system, we have applied a simple tight-binding (TB)
model~\cite{TB} to fit the present LAPW results at six symmetry points
in the BZ. This TB model has included the V 3$d$ as well as the O 2$s$
and 2$p$ orbitals. Using 12 independent TB parameters, a moderately
accurate (rms error = 0.17 eV) fit has been obtained to the 52
LiV$_2$O$_4$ valence and conduction bands. The fitting has involved
two-center energy as well as overlap parameters. The TB parameters are
listed in Table I. The TB representation of the LiV$_2$O$_4$ bands is
shown in Fig.~\ref{band}(b). It is clear that, although the agreement
is not perfect, the TB model captures the essential features of the
LAPW results.

A different TB model excluding the oxygen orbitals has been applied to
the LAPW V 3$d$ band results in order to obtain estimates of the
orbital energies for the individual $a_{1g}$ and $e_{g}$ subbands in
this system. (Note that the TB parameters in Table I include only a
single V 3$d$ orbital energy $E_d$). These TB orbital energies
represent the mean band energy for each subband. They are in fact the
crystal-field levels that come into play in the limit where the V 3$d$
electrons are localized by electron-correlation effects. According to
the present analysis, the mean band energy for the upper $e_{g}$
complex is 2.82 eV while those for the $t_{2g}$-derived $a_{1g}$ and
$e_{g}$ manifold are 0.43 and 0.29 eV, respectively. Thus,
crystal-field splitting that arises from the octahedral coordination
of oxygens (2.48 eV) is more than an order-of-magnitude larger than
that originating from the trigonal distortion ($\sim$0.14 eV). This TB
model, which also involved a total of 12 parameters, yielded a
moderate fit (rms error = 0.15 eV) to the V 3$d$-band states. The
effective $d$-$d$ hopping integrals [$(dd\sigma)$, $(dd\pi)$,
$(dd\delta)$] over three shells of V-V neighbors [$d_1$ = 2.91 \AA,
$d_2$ = 5.04 \AA, $d_3$ = 5.82 \AA] that have been included in this
fit have values (in eV) [-0.425, 0.008, 0.152], [-0.034, 0.026,
-0.021], and [-0.060, 0.064, -0.026], respectively. These parameters
may provide a useful starting point for future studies of
electron-electron correlation effects in this system.
 
\begin{figure} 
\epsfxsize=80mm \centerline{\epsfbox{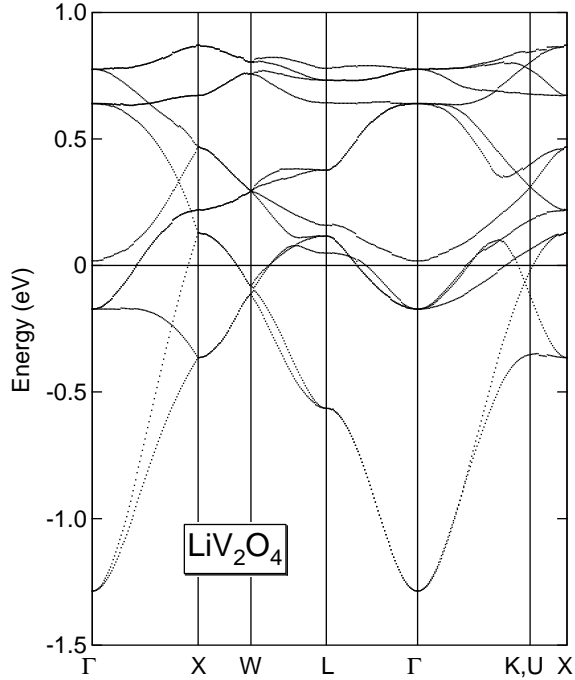}} \vspace{0.5pc}
\caption{Expanded plot of the LAPW $t_{2g}$ bands for LiV$_2$O$_4$.}
\label{band_expanded} 
\end{figure}

A more detailed view of the $t_{2g}$ portion of the LAPW V 3$d$
energy-band results for LiV$_2$O$_4$ is shown in
Fig.~\ref{band_expanded}. The six LiV$_2$O$_4$ valence electrons per
cell are sufficient to fill, on average, three of the 12 $t_{2g}$
conduction bands. In fact, the calculated LAPW band dispersion
produces several partially-filled bands, leading to a rather
complicated Fermi-surface topology in this system. While two bands are
completely filled, the third conduction band contains holes near the
$X$ and $L$ symmetry points ($h_3$) and electron pockets at $\Gamma$
($e_4$ and $e_5$) and $W$ ($e_4$). Since the unit cell contains an
even number of electrons, compensation requires that there exist equal
numbers (i.e. BZ volumes) of electrons and holes. The existence of
Fermi-surface sheets with both electron and hole character in
LiV$_2$O$_4$ provides a ready explanation for the experimental
observation that the Hall coefficient changes sign with
temperature~\cite{Takagi} in accordance with a two-carrier model.

\begin{figure}
\epsfxsize=80mm \centerline{\epsfbox{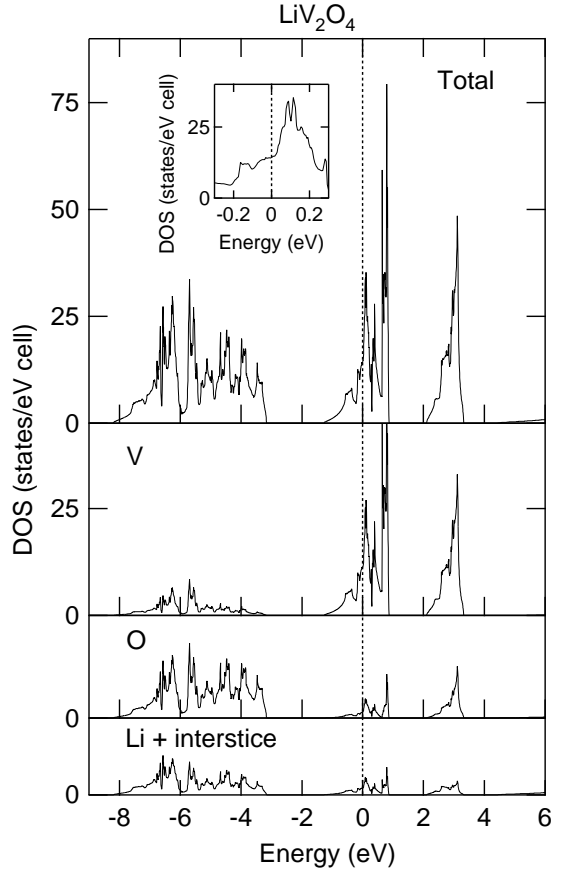}} \vspace{0.5pc}
\caption{Total and muffin-tin projected DOS results for
 LiV$_2$O$_4$. The inset shows an enlarged view of the total DOS near
 $E_F$.}
\label{dos} 
\end{figure}

The total DOS and the projected contributions to the DOS from various
muffin-tins are shown in Fig.~\ref{dos}. The dominant contribution to
the DOS over this energy range originates from the O 2$p$ and V 3$d$
states. The O 2$p$ bands are filled by electrons transferred from the
Li 2$s$ and V 3$d$ orbitals so the system can be described by the
ionic configuration Li$^+$(V$^{3.5+}$)$_2$(O$^{2-}$)$_4$. Previous
photoemission-spectra data are in general agreement with these
results.~\cite{Fujimori} However, the calculated band DOS and the
photoemission spectra do not agree with each other in the V 3$d$ band
region near $E_F$. In particular, the photoemission intensity at $E_F$
is an order of magnitude lower than the calculated DOS. On the other
hand, the density of quasi-particles at $E_F$ derived from the
specific heat is much higher than the calculated $D(E_F)$. According
to our results, the calculated $D(E_F)=$ 7.1
states/(eV$\cdot$formula-unit) corresponds to $\gamma_{\rm cal} =$ 17
mJ/mol K$^2$; this is $\sim$25 times smaller than the experimental
value of $\gamma \approx$ 0.42 J/mol K$^2$. It is well-known that
electron-phonon interactions enhance the electronic specific heat as
in the closely-related spinel compound LiTi$_2$O$_4$, where band
calculations indicated strong electron-phonon-coupling effects,
yielding a coupling parameter $\lambda=$ 1.8 and hence a mass
enhancement factor of $1+\lambda=$ 2.8.~\cite{Satpathy} In the present
case, however, the deduced enhancement factor of $\sim$25 suggests
that a different mechanism such as electron-correlation effects would
be responsible for the observed HF behavior in LiV$_2$O$_4$.

In comparison with the band structures of typical 4$f$-electron HF
systems, our result show similar behavior in that the calculated
density of states at $E_F$ is much smaller than those obtained from
the analysis of experimental specific-heat data. For example, the
ratio $\gamma_{\rm exp}/\gamma_{\rm cal}$ is $\sim$100 for
CeCu$_2$Si$_2$ and $\sim$70 for CeAl$_3$.~\cite{CeCu2Si2} Also, the
band structures of the $f$-electron HF and LiV$_2$O$_4$ systems are
similar in that there is a sharp DOS peak just above $E_F$ ($\sim$0.3
eV for CeCu$_2$Si$_2$ and $\sim$0.1 eV for LiV$_2$O$_4$, see inset of
Fig.~\ref{dos}). In order to explain the high density of
quasi-particles deduced from the electronic specific heats, a
renormalized band picture was proposed, in which a strong
renormalization of the $f$ band due to the strong correlation at the
4$f$ site was taken into account.~\cite{renormalization} In the
renormalized band picture, it is considered that electrons with two
different degrees of wave function localization, namely, the itinerant
$sd$ conduction electrons and the well localized $f$ electrons
hybridize with each other. In the case of LiV$_2$O$_4$, however, all
the conduction bands crossing $E_F$ consist of V $t_{2g}(3d)$
orbitals, i.e., electrons with the same degrees of localization, and
therefore the mechanism of the mass enhancement may be totally
different from the $f$-electron systems. The narrowing of the energy
bands due to electron correlation within the $t_{2g}$ bands is likely.
Given a high-temperature local moment behavior in a metallic system, a
prerequisite for a HF behavior is that the local moments do not show
long-range order at low temperatures. In the 4$f$-electron HF systems,
long-range order is prohibited by weakness of the coupling between the
4$f$ local moments.~\cite{Yamada} In the case of LiV$_2$O$_4$, the
long-range order is disfavored by the magnetic frustrations
originating from the spinel structure. This phenomenological
consideration remains to be justified by microscopic theories.

In conclusion, the results of LAPW band calculations for the
spinel-type compound LiV$_2$O$_4$ show that the O 2$p$, V 3$d$
($t_{2g}$) and V 3$d$ ($e_{g}$) bands as well as the higher-lying $sp$
conduction band are well separated from each other and that the Fermi
level lies within the $t_{2g}$ bands. The electronic structure of
LiV$_2$O$_4$ can be well described by electrons in the
triply-degenerate $t_{2g}$ bands, thereby indicating that the
mechanism of the mass enhancement in this compound should be different
from that in the $f$-electron heavy Fermion systems. Further
experimental studies on various physical properties at low and high
temperatures are necessary to characterize the nature of the heavy
Fermion behavior in LiV$_2$O$_4$.  High-resolution photoemission
studies would be especially useful for clarifying the renormalization
of quasi-particles in this system.

The authors would like to thank Prof.~Y.~Ueda for valuable
discussions. This work was supported by a Special Coordination Fund
for Promoting Science and Technology from the Science and Technology
Agency of Japan. L.~F.~M. is grateful to Yamada Science Foundation for
supporting his visit to the University of Tokyo. J.~M. acknowledges
support from the Japan Society for the Promotion of Science for Young
Scientists.

\newpage
\mediumtext
\begin{table}[h]
\caption{Tight-binding parameters for LiV$_2$O$_4$ (rms error = 0.17 eV).
Parameters in braces \{...\} have been approximated using simplifying 
relations (i.e., ($dd\pi$) = 0.5$|(dd\sigma)|$, etc.), leaving 12 independent
parameters.}
\begin{tabular}{ccll}
Site/interaction&Distance (\AA) &Parameters  &Value (eV)\\
\tableline
  & & Energy &\\
 V &...& $E_d$   & $-$0.569 \\
 O &...& $E_s$, $E_p$ & $-$18.715, $-$4.390 \\
 V-V&2.91& $(dd\sigma)$, $(dd\pi)$, $(dd\delta)$&$-$0.524, \{0.262\}, \{$-$0.052\}\\
 V-O&1.97& $(sd\sigma)$, $(pd\sigma)$, $(pd\pi)$&$-$2.278, $-$2.190, \{1.095\}\\
 O-O&2.65& $(ss\sigma)$, $(sp\sigma)$, $(pp\sigma)$, $(pp\pi)$& $-$0.239,
 \{0.413\}, 0.711, \{$-$0.284\}\\
 O-O&2.91& $(ss\sigma)$, $(sp\sigma)$, $(pp\sigma)$, $(pp\pi)$&
 $-$0.110, \{0.229\}, 0.479, \{$-$0.192\}\\
  & & Overlap &\\
 V-O&1.97& $[sd\sigma]$, $[pd\sigma]$, $[pd\pi]$ & 0.046, 0.051, \{$-$0.026\}
\end{tabular}
\end{table}
\end{document}